\documentclass[10pt,letterpaper]{article}
\usepackage{opex3,amsmath,amssymb,graphicx,cancel}
\usepackage{cite}

\begin{document}

\title{On the dynamics of Airy beams in nonlinear media with nonlinear losses}

\author{Carlos Ruiz-Jim\'enez,$^{1}$ K. Z. N\'obrega,$^2$ and Miguel A. Porras$^{3,*}$}

\address{$^1$Grupo Sistemas Complejos, Universidad Polit\'ecnica de Madrid, Ciudad Universitaria s/n, 28040 Madrid, Spain\\
$^2$Federal Institute of Maranh\~ao, Department of Electro-Electronics, S\~ao Luis-Ma, Brazil\\
$^3$Grupo de Sistemas Complejos, Universidad Polit\'{e}cnica de Madrid, Rios Rosas 21, 28003 Madrid, Spain}

\email{$^*$miguelangel.porras@upm.es}

\begin{abstract}
We investigate on the nonlinear dynamics of Airy beams in a regime where nonlinear losses due to multi-photon absorption are significant. We identify the nonlinear Airy beam (NAB) that preserves the amplitude of the inward H\"ankel component as an attractor of the dynamics.
This attractor governs also the dynamics of finite-power (apodized) Airy beams, irrespective of the location of the entrance plane in the medium with respect to the Airy waist plane. A soft (linear) input long before the waist, however, strongly speeds up NAB formation and its persistence as a quasi-stationary beam in comparison to an abrupt input at the Airy waist plane, and promotes the formation of a new type of highly dissipative, fully nonlinear Airy beam not described so far.
\end{abstract}


\section{Introduction}

The discovery by Berry and Balazs \cite{BERRY} in Quantum Mechanics of the existence of non-spreading, self-accelerating Airy wave packets hibernated for decades in the literature until its experimental realization in the form of an Airy beam of light \cite{SIVI,SIVI2}. It was indeed in Optics where the land was prepared thanks to the re-discovery in the eighties of non-diffracting Bessel beams \cite{DURNIN}. Since then huge amount of work has been made on self-accelerating Airy beams, including their properties \cite{BANDRES,BROKY}, propagation dynamics in different nonlinear media \cite{ABDOLLAHPOUR,PANAGIOTOPOULOS,FATTAL,YIQI,LOTTI,PANAGIOTOPOULOS2}, nonparaxial \cite{PENG}, vectorial \cite{KAMINER2} and fully nonlinear \cite{KAMINER} generalizations, and applications as diverse as particle clearing \cite{BAUNGARTL} or creation of curved plasma channels \cite{POLYNKIN}. It seems that this amount of work was necessary to close the circle with the recent creation of genuine quantum Airy wave packets of free electrons \cite{VOLOCH-BLOCH}.

In this paper we are interested in the nonlinear propagation of one-dimensional Airy beams in a regime where nonlinear losses (NLLs) due to multiphoton absorption are significant. In this regime, Lotti {\em et al} \cite{LOTTI} have demonstrated theoretically the existence of Airy-like beams that can propagate stationarily, overcoming the attenuating effects of NLLs. These nonlinear Airy beams (NABs) cannot have arbitrary intensities, but their peak intensity is always lower than a maximum value determined by the optical properties of medium. The mechanism of stationarity of NABs is similar to that described earlier for nonlinear Bessel beams in the same regime \cite{PORRAS3,PORRAS}. Asymptotically, a NAB behaves as a linear Airy beam except that the amplitudes, say $|a_{\rm in}|$ and $|a_{\rm out}|$, of its inward and outward H\"ankel components are not equal. The excess of the inward H\"ankel component creates an inward power flux that refills the power absorbed mainly in the more prominent, central intensity maxima. NABs are characterized by a loss of contrast and compression (in self-focusing Kerr media) of the intensity ripples in their intensity profile compared to those of the linear Airy beam \cite{LOTTI}. The numerical simulations and experiments in \cite{LOTTI} suggest that a linear Airy beam introduced in a nonlinear medium with NLLs transforms into a stationary NAB because the loss of contrast and compression characteristic of NABs were indeed observed. Also, more detailed numerical simulations in \cite{COUAIRON} evidence the trend of the input linear Airy beam towards a stationary NAB. In case that a stationary NAB is (or tends to be) formed, a relevant question is which particular one, among all supported by the medium, is selected as the final stationary state for a given input Airy beam, so that its intensity can be predicted in practice. Further, finite-power Airy beams, e. g., exponential-Airy beams, have a transversal intensity profile changing from Gaussian at the far field to exponential-Airy at the ``waist" \cite{SIVI}. Another relevant question is then whether these different entrance conditions of the input Airy beam may lead to NAB formation or not, and in the affirmative case, whether the formed NABs are different or not.

Our numerical simulations with different values of the intensity, apodization parameter and entrance plane of the input Airy beam shows in fact seemingly disparate behaviors. We show however that all them can be understood from a simple conservation principle. When an Airy beam ($a_{\rm in}=a_{\rm out}\equiv a_0$), apodized or not, at its waist or not, is launched into the nonlinear medium, the NAB that preserves the amplitude of the asymptotic inward H\"ankel component, i. e., that with $|a_{\rm in}|=a_0$, tends to be spontaneously formed. A similar conservation principle has been given recently for Bessel beams \cite{PORRAS}. The nonlinear dynamics of Airy beams is then determined by the asymptotic behavior of NABs, which is analyzed in detail in Sec. \ref{ASYMP}.

With ideal Airy beams, the stationary attracting NAB is fully formed, as seen in Sec. \ref{SELECTION}. With finite-power Airy beams, the situation is more complex, through the trend towards the same attracting NAB continues to dominate the dynamics, as seen in Sec. \ref{APODIZED}. As two representative cases with seemingly opposite behavior, we consider exponentially apodized Airy beams entering the medium at their waist, where they are fully formed and more intense (nonlinear or ``abrupt" input), and entering the medium long before the waist, where they are widespread and Gaussian-like, so that nonlinearities are initially negligible (linear or ``soft" input). It is noteworthy that the same NAB tends to be formed under such a disparate input conditions. With abrupt input, however, we only observe the first steps, mimicking the ideal case, towards the formation of the attracting NAB, but the beam decays well before it can be reached. Actually, no vestige of stationary behavior can be identified. With soft input of the same Airy beam, the attracting NAB is found to be much more closely reached. Its transversal profile is clearly observable and is seen to behave as a quasi-stationary light beam.

Also, we identify a propagation regime in the limit of very high power of the input linear Airy beam for which NABs with asymptotic H\"ankel tail are no longer formed. What is observed is instead a finite-power version of the limiting NAB (LNAB) supported by the medium. This LNAB does not have a linear H\"ankel tail, but is nonlinear everywhere, strongly dissipative (infinitely dissipative, ideally) and more weakly localized than NABs. This attracting LNAB in this limit of high powers is solely determined by the optical properties of the medium.

\section{Asymptotic behavior of ideal nonlinear Airy beams}\label{ASYMP}

Our starting point is the nonlinear Schr\"odinger equation (NLSE)
\begin{equation}\label{NLSE}
\partial_z A =\frac{i}{2k}\partial^2_x A + i\frac{kn_2}{n}|A|^2A - \frac{\beta^{(M)}}{2}|A|^{2M-2}A \,,
\end{equation}
for the complex envelope $A(x,z)$ of the light beam $E=A\exp[-i(\omega t-kz)]$ of angular frequency $\omega$ and linear propagation constant $k=(\omega/c)n$. In the above equations $n$ is the linear refractive index, $c$ the speed of light in vacuum, $n_2$ is the nonlinear refractive index, and $\beta^{(M)}>0$ is the $M$-photon absorption coefficient responsible for NLLs. For a more comprehensive analysis we introduce the dimensionless coordinates $\xi=x/x_0$, $\zeta=z/(kx_0^2)$, where $x_0$ is an arbitrary transversal scale, the dimensionless envelope $\tilde A=(kx_0^2\beta^{(M)}/2)^{1/(2M-2)}A$, and describe the evolution of the envelope in the uniformly accelerated reference frame defined by $u=\xi-(\zeta/2)^2$, $v=\zeta$. With these changes Eq. (\ref{NLSE}) becomes
\begin{equation}\label{NLSEN1}
\partial_v \tilde A =\frac{i}{2}\partial^2_u \tilde A +\frac{v}{2}\partial_u \tilde A+ i\alpha|\tilde A|^2\tilde A - |\tilde A|^{2M-2}\tilde A \,,
\end{equation}
where $\alpha=(k^2x_0^2n_2/n)(2/kx_0^2\beta^{(M)})^{(1/M-1)}$. In the absence of the nonlinear terms, Eq. (\ref{NLSEN1}) admits the linear Airy beam solution \cite{BERRY} $\tilde A(u,v) \propto Ai(u)\exp[i\phi_L(u,v)]$, where $\phi_L(u,v)=uv/2+v^3/24$, whose intensity profile is invariant in the accelerated reference system. With the nonlinear terms, we search, as in \cite{LOTTI}, for stationary solutions of the form $\tilde A=a(u)\exp[i\psi(u)]\exp[i\phi_L(u,v)]$, where  $a(u)$ is the real amplitude and $\psi(u)$ is a nonlinear phase. The NLSE in Eq. (\ref{NLSEN1}) then yields
\begin{eqnarray}\label{NLEQS}
a^{\prime\prime}&=&ua+(\psi^{\prime})^2a-2\alpha a^3, \nonumber \\
\psi^{\prime\prime}&=&-2\frac{\psi^\prime a^\prime}{a}-2a^{2M-2},
\end{eqnarray}
where primes stand for $d/du$. The second equation is readily seen to be equivalent to
\begin{equation}\label{NLL}
\psi^\prime a^2=2\int_u^\infty a^{2M}du\equiv N_u,
\end{equation}
which is the refilling condition for stationarity with NLLs, stating that the NLLs $N_u$ in $[u,\infty)$ must be compensated by a positive flux $\psi^\prime a^2$ at $u$. NABs are solutions of these equations behaving as $a(u)\rightarrow a_0Ai(u)$, $\psi(u)\rightarrow 0$, i. e., as the linear Airy beam, for $u\rightarrow +\infty$, and verifying the localization condition $a(u)\rightarrow 0$ for $u\rightarrow-\infty$. An asymptotic analysis reveals several possibilities for the decay for $u\rightarrow-\infty$, depending on the admissible values of the total NLLs $N_{-\infty}$ that the beam experience per unit propagation length.

\subsection{Nonlinear Airy beams with finite nonlinear losses}\label{ASYMPFINITE}

As in \cite{LOTTI}, we first assume that $N_{-\infty}<\infty$. Replacing $N_{u}$ with $N_{-\infty}$ in Eq. (\ref{NLL}) at large enough negative $u$, i. e., $\psi'a^2\backsimeq N_{-\infty}$, and with the change $a=b(\rho)/(-u)^{1/4}$, where $\rho=(2/3)(-u)^{3/2}$, we derive the Newton equation $d^2b/d\rho^2=-b+N_{-\infty}^2/b^3$ (after neglecting subleading terms in $\rho$), which is the equation of motion of a particle subjected to a conservative force with potential $V=b^2/2+1/2N_{-\infty}^2/b^2$, and therefore with a conserved ``energy" $E=(1/2)(db/d\rho)^2+V$. The general solution of the Newton equation leads to the possible asymptotic behavior for the intensity
\begin{equation}\label{NABASYM1}
a^2\backsimeq \frac{E}{(-u)^{1/2}}\left\{1+C\,\sin\left[\frac{4}{3}(-u)^{3/2}+\phi\right]\right\}\, ,
\end{equation}
where $C\equiv (1-N_{-\infty}^2/E^2)^{1/2}$ and $N_{-\infty}<E$. This is the asymptotic behavior of the NABs reported in \cite{LOTTI}, decaying as the linear Airy beam for $u\rightarrow -\infty$ but with possibly shifted oscillations by a phase $\phi$ and reduced contrast $C<1$. These NABs are therefore asymptotically linear waves, whose NLLs are indeed finite and concentrated in the nonlinear core close to $u=0$. Considering the asymptotic expressions of H\"ankel functions of order $1/3$ for large negative argument \cite{OLVER}, and as pointed out in \cite{LOTTI}, the above asymptotic behavior of the intensity $a^2$ and the flux $\psi' a^2\backsimeq N_{-\infty}$ is reproduced by the expression \cite{LOTTI2}
\begin{equation}\label{NABASYM2}
\tilde A\backsimeq\frac{1}{2}\sqrt{\frac{-u}{3}}\left[a_{\rm out}{\rm e}^{i\pi/6}H_{1/3}^{(1)}(\rho)+a_{\rm in}{\rm e}^{-i\pi/6}H_{1/3}^{(2)}(\rho)\right],
\end{equation}
provided that $|a_{\rm out}|^2=2\pi(E-N_{-\infty})$, $|a_{\rm in}|^2=2\pi(E+N_{-\infty})$, and $\mbox{arg} (a_{\rm out})-\mbox{arg} (a_{\rm in})=\phi$.  The linear Airy beam $\tilde A = a_0Ai(u)\exp[i\phi_L(u,v)]$ is recovered with $a_{\rm in}= a_{\rm out}=a_0$. For NABs $|a_{\rm in}|^2>|a_{\rm out}|^2$, indeed $|a_{\rm in}|^2-|a_{\rm out}|^2= 4\pi N_{-\infty}>0$. Using the relations between H\"ankel and Airy functions of the first and second type \cite{OLVER}, Eq. (\ref{NABASYM2}) can be interestingly written in more compact form as $\tilde A\backsimeq pAi(u)+iqBi(u)$,
where $p=(a_{\rm in}+a_{\rm out})/2$ and $q=(a_{\rm in}-a_{\rm out})/2$.

\begin{figure}[t]
\includegraphics*[width=6.6cm]{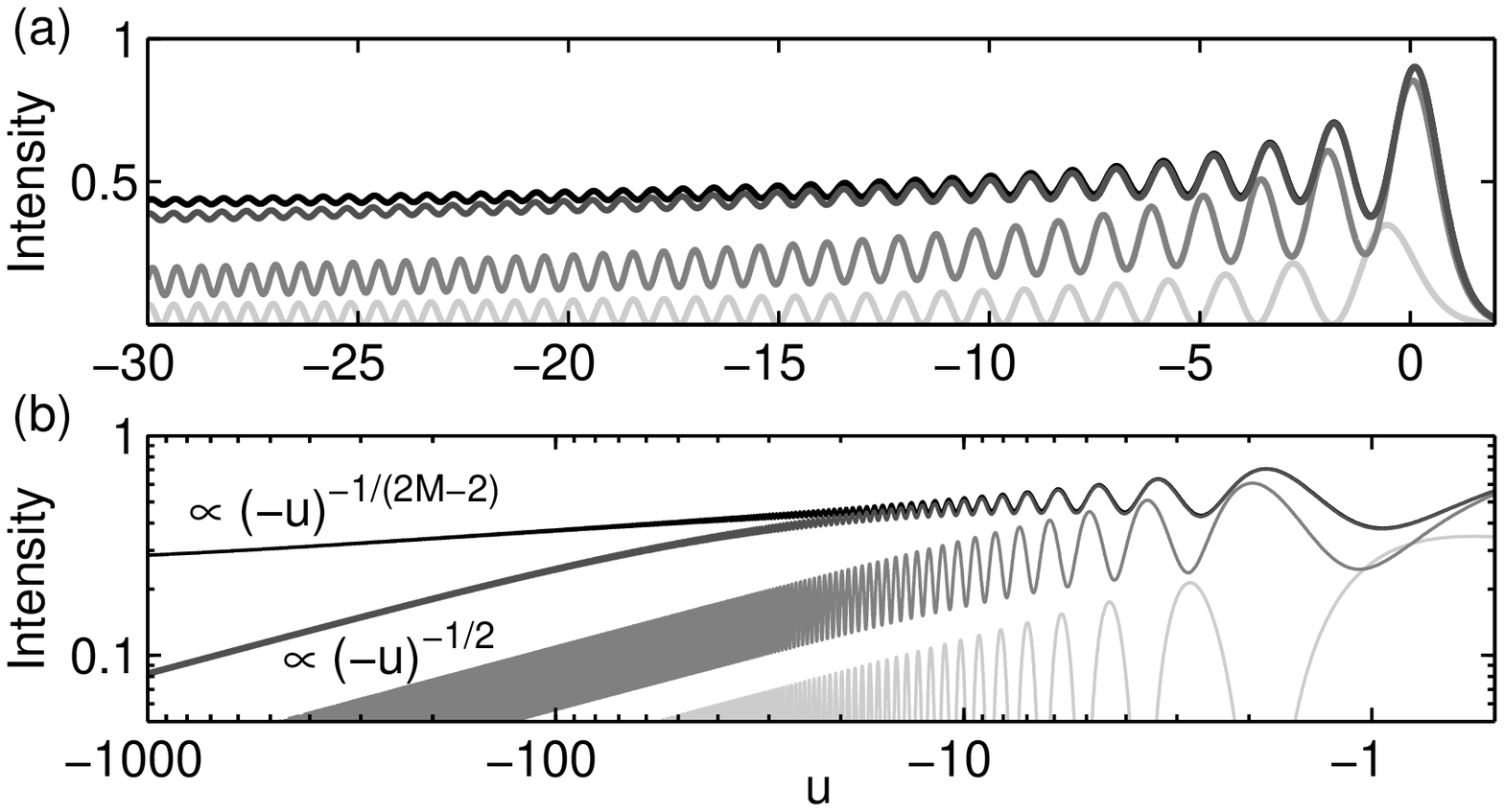}
\includegraphics*[width=3.5cm]{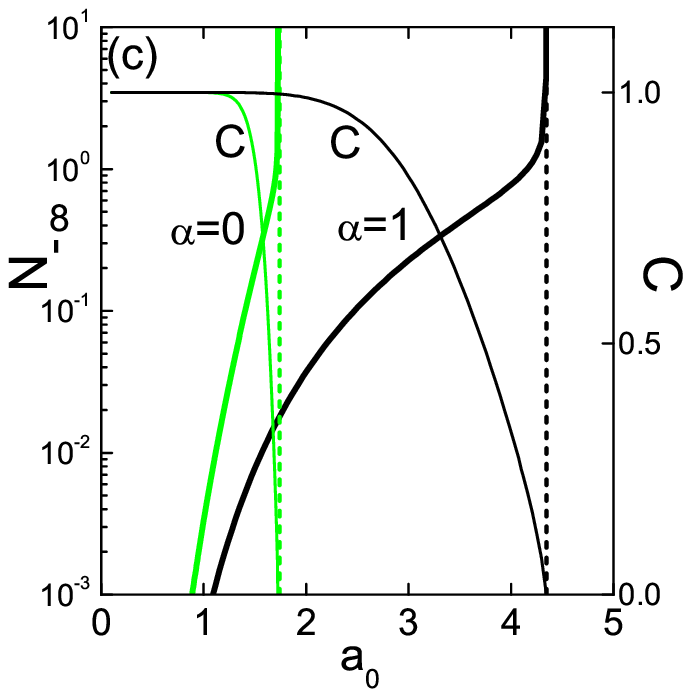}\includegraphics*[width=3.3cm]{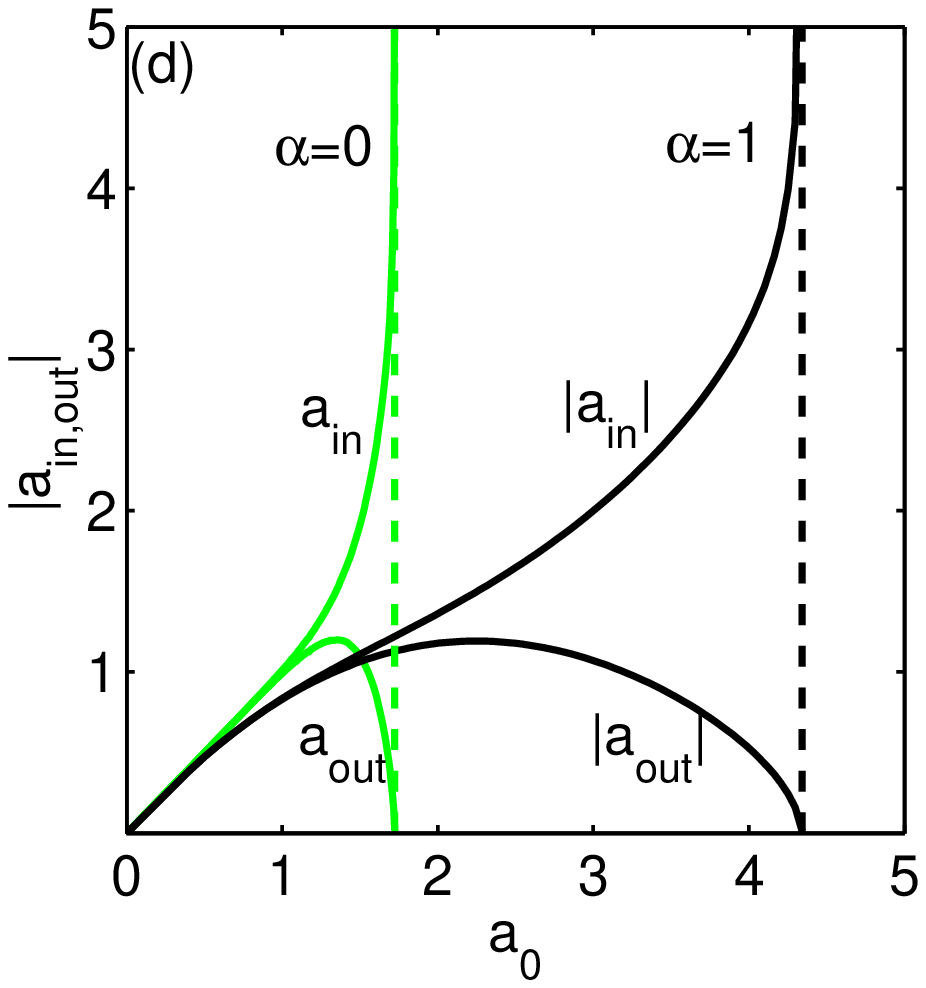}
\caption{\label{Fig1}
(a) Gray curves: Intensities profiles of NABs with $a_0=1.500$, $a_0=4.000$ and $a_0=4.325$ in a medium with $M=5$ and Kerr nonlinearity $\alpha=1$. Black curve: Intensity profile of the LNAB with $a_0=a_{0,\rm max}\simeq 4.343$. (b) The same but in double logarithmic scale. The curves decay as $(-u)^{-1/2}$ for the NABs and as $(-u)^{-1/(2M-2)}$ for the LNAB for $u\rightarrow-\infty$. (c) Nonlinear losses $N_{-\infty}$ and oscillation contrast $C\equiv (1-N_{-\infty}^2/E^2)^{1/2}$ of NABs as functions of $a_0$. (d) Amplitudes $|a_{\rm in}|$ and $|a_{\rm out}|$ of the inward and outward H\"ankel components of NABs as functions of $a_0$. In (b) and (c), $M=5$, and $\alpha=0$ (green curves), and $\alpha=1$ (black curves).}
\end{figure}

\begin{figure}
\center\includegraphics*[width=8cm]{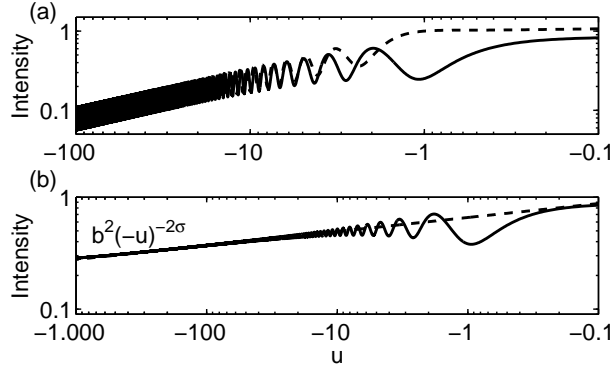}
\caption{\label{Fig2}
In a medium with $M=5$ and Kerr nonlinearity $\alpha=1$, in double logarithmic scale, (a) intensity profile of the NAB with $a_0=4.0$ (solid curve), and its asymptotic form in Eq. (\ref{NABASYM2}) with $|a_{\rm in}|=3.176$ and $|a_{\rm out}|=0.523$ (dashed curve); (b) intensity profile of the LNAB with $a_{0,\rm max}=4.343$ (solid curve) and its asymptotic form $a^2\sim b^2 (-u)^{-2\sigma}$ with $b$ and $\sigma$ in Eq. (\ref{LNABASYM1}) (dashed curve).}
\end{figure}

Numerical integration of Eqs. (\ref{NLEQS}) with initial condition $a(u_0) = a_0Ai(u_0)$ and $\psi=0$ at some large positive $u_0$, shows that NABs with the above asymptotic behavior exist only for $a_0<a_{0,\rm max}$, where $a_{0,\rm max}$ is a number of the order of unity that depends on $M$ and the strength of Kerr nonlinearity $\alpha$ (e. g., for $M=5$, $a_{0,\rm max}\simeq 1.72$ for $\alpha=0$ and $a_{0,\rm max}\simeq 4.34$ for $\alpha=1$). The three gray curves in Fig. \ref{Fig1}(a) represent the intensity profiles of NABs with increasing $a_0$ in a medium with $M=5$ and $\alpha=1$. As $a_0$ increases, the peak intensity increases at first, but saturates as $a_0$ approaches $a_{0,\rm max}$, showing that NABs can exist only up to a maximum value of the peak intensity. The same NABs are represented in Fig. \ref{Fig1}(b) in double logarithmic scale in order to appreciate their asymptotic decay as $1/(-u)^{1/2}$. Figure \ref{Fig1}(c) shows that the total NLLs $N_{-\infty}$ of NABs are finite for any $a_0$, but grow without limit as $a_0\rightarrow a_{0,\rm max}$, and that the contrast of the oscillations decreases down to zero in this limit. The values of $N_{-\infty}$ and $C$ allow us to find the values of the amplitudes $|a_{\rm in}|$ and $|a_{\rm out}|$ of the inward and outward H\"ankel components, which are represented as functions of $a_0$ in Fig. \ref{Fig1}(d). These curves turn out to be essential for understanding the nonlinear dynamics of Airy beams, either ideal or their finite-energy realizations. In absence of Kerr nonlinearity ($\alpha=0$), $a_{\rm in}$ and $a_{\rm out}$ turn out numerically to be real, while with dispersive nonlinearities as Kerr nonlinearity $\mbox{arg}(a_{\rm out})=-\mbox{arg}(a_{\rm in})$.
Similar asymptotic analysis has been made recently for vortex Bessel beams \cite{PORRAS}. A particular NAB and its asymptotic form in Eq. (\ref{NABASYM2}) can be seen in Fig. \ref{Fig2}(a).

\subsection{Nonlinear Airy beams with infinite nonlinear losses}\label{ASYMPINFINITE}

We point out that finiteness of $N_{-\infty}$ is a reasonable but not necessary assumption. Localized wave objects with infinite power and infinite nonlinear losses have been previously described and have been shown to play a role in actual physical processes, as in the self-focusing of Gaussian wave packets halted by nonlinear losses \cite{PORRAS2}.

Let us assume a more general decay $a\sim b (-u)^{-\sigma}$ for $u\rightarrow -\infty$, where $b$ and $\sigma$ are positive constants, and that $\sigma$ is small enough so that $N_u\rightarrow \infty$ for $u\rightarrow -\infty$. Integrating Eq. (\ref{NLL}), we get $\psi^\prime \backsimeq 2b^{2M-2}(-u)^{1-2\sigma(M-1)}/(1-2M\sigma)$, where we have neglected a finite constant compared to the assumed infinitely large NLLs.
Introducing this expression into the first of Eqs. (\ref{NLEQS}) we obtain
\begin{equation}
b\sigma(\sigma\!+\!1)(-u)^{-\sigma-2}+b(-u)^{-\sigma+1}-\frac{4b^{4M-3}}{(1\!-\!2M\sigma)^2}(-u)^{2-\sigma(4M-3)} +2\alpha b^3(-u)^{-3\sigma}\backsimeq 0.
\end{equation}
for $u\rightarrow -\infty$. The second and third terms in the left hand side cancel each other asymptotically, and the other two terms are negligibly small, if
\begin{equation}\label{LNABASYM1}
b =\left[\frac{M-2}{4(M-1)}\right]^{1/(2M-2)}, \quad \sigma =\frac{1}{4M-4} .
\end{equation}
With this value of $\sigma$, $N_{-\infty}$ tends indeed to infinity.

From a purely numerical approach, it turns out that the transversal profiles of NABs with $a_0$ approaching $a_{0,\rm max}$ [the upper gray curve in Figs. \ref{Fig1}(a) and \ref{Fig1}(b)] tend to match a certain transversal profile [the black curve in Figs. \ref{Fig1}(a) and \ref{Fig1}(b)], whose behavior is described by Eqs. (\ref{LNABASYM1}). The matching region expands toward negative values of $u$ as $a_0$ increases up to the limiting value $a_{0,\rm max}$, and fills the whole space down to $u=-\infty$ in the limit $a_0=a_{0,\rm max}$. The asymptotic Airy-like behavior of NABs, and in particular its structure as two interfering H\"ankel beams, thus disappears in this limit, and is replaced by the asymptotic behavior of this limiting nonlinear Airy beam (LNAB) with $a_0=a_{0,\rm max}$. Contrary to NABs, NLLs are not concentrated in the nonlinear core, but are significant in the whole transversal profile. We note that this LNAB has no free parameters, but is solely determined by the properties of the medium ($M$ and $\alpha$ in our dimensionless variables). Figure \ref{Fig2}(b) shows the LNAB in a medium with $M=5$ and $\alpha=1$ and its asymptotic behavior determined by Eq. (\ref{LNABASYM1}). Despite the ideal behavior of the LNAB, it will seen to act as an attractor in the nonlinear propagation of very powerful, real Airy beams. In another context, fully nonlinear and dissipative solitons in media with spatially-inhomogeneous nonlinear absorption, but requiring linear gain, have been reported in \cite{BOROVKOVA}.

\section{Dynamics of ideal Airy beams in the nonlinear medium} \label{SELECTION}

We turn now into the problem of the propagation of Airy beams in nonlinear media. We are interested in an intensity regime where the effects of NLLs  are significant, being comparable to, or even dominant over other nonlinear effects (Kerr nonlinearity in our case). This means in particular that the possible solitary structures emerging from the Airy beam due to the Kerr nonlinearity, as described in \cite{YIQI}, are significantly affected by NLLs. This condition is fulfilled if the intensity $n_0/(k_0^2n_2x_0^2)$ required for soliton formation is higher than, or of the order of the typical intensity $(2/\beta^{(M)}k x_0^2)^{1/(M-1)}$ at which NLLs are significant. In our dimensionless variables, this means that $\alpha$ does not exceed unity significantly.

\begin{figure}
\centering\includegraphics*[width=4.5cm]{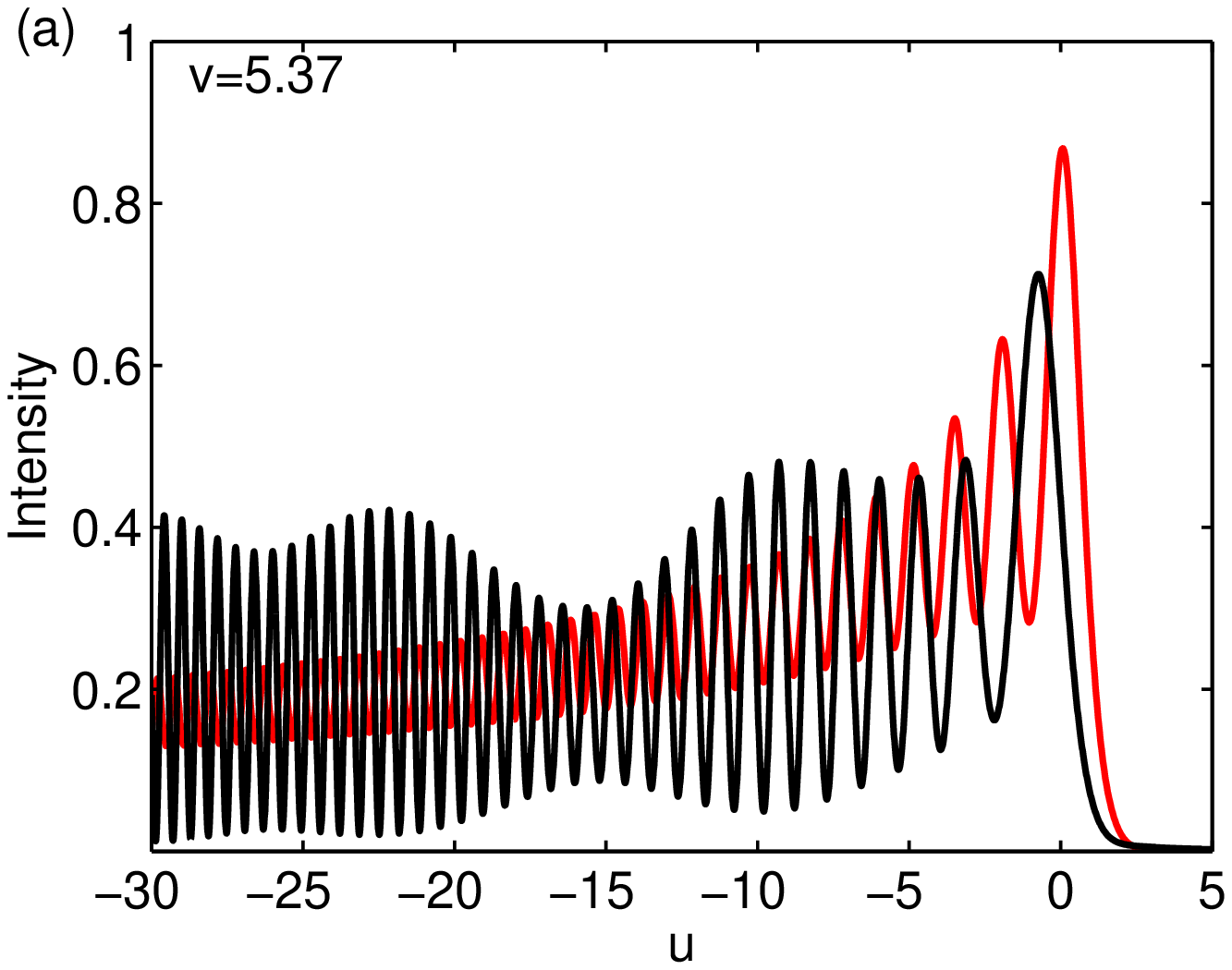}\includegraphics*[width=3.6cm]{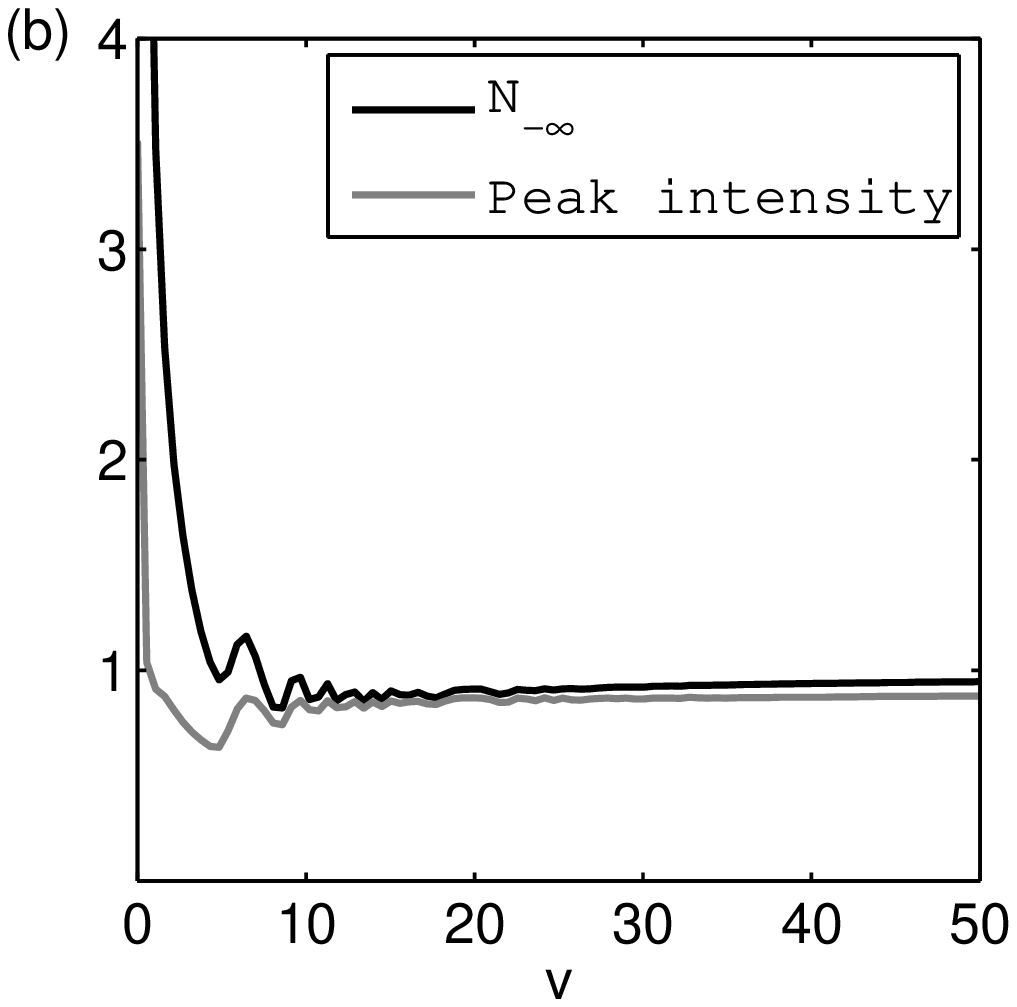}\includegraphics*[width=3.6cm]{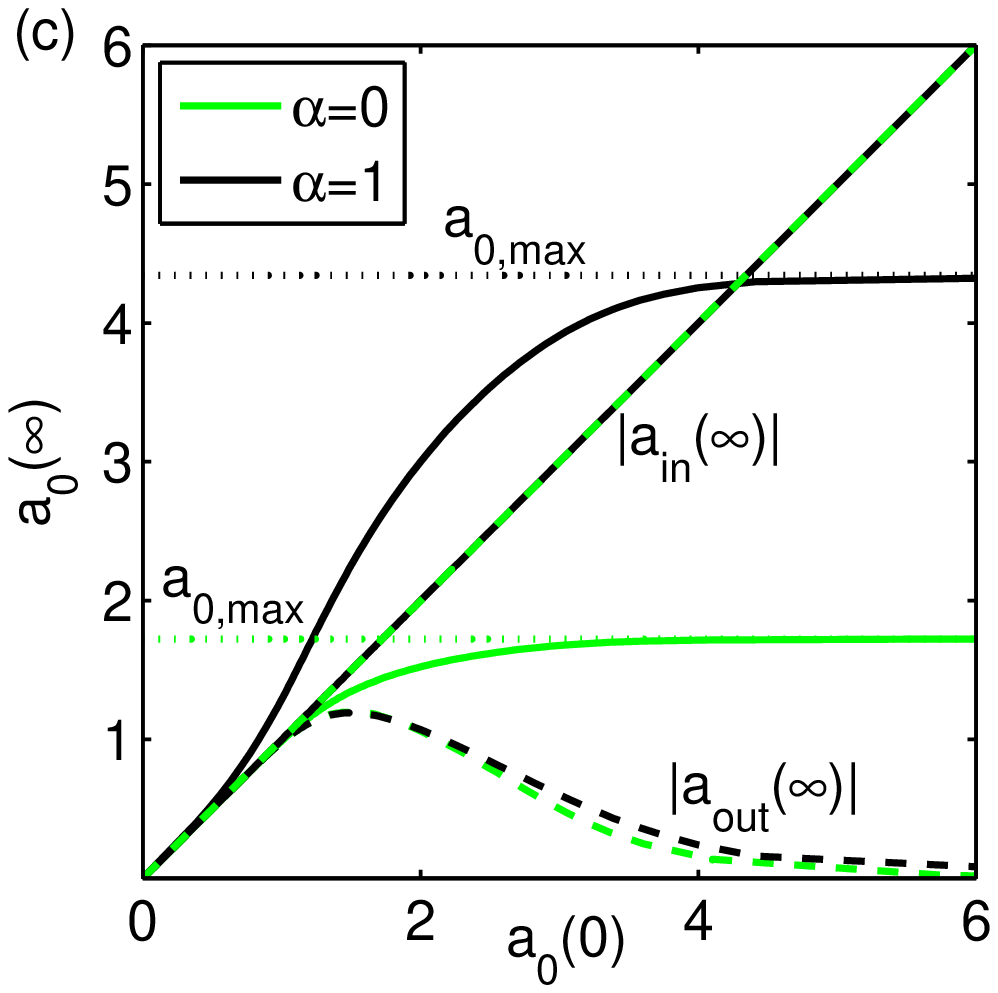}
\caption{\label{Fig3} For a medium with $M=5$ and $\alpha=1$, (a) \textcolor{blue}{Media˜1}: Change of the intensity profile of an input Airy beam with $a_0=3.5$ with propagation distance $v$. The red curve represents the final attracting NAB with $a_0=4.13$, whose inward H\"ankel amplitude is $|a_{\rm in}|=3.5$. (b) NLLs $N_{-\infty}=2\int_{-\infty}^{\infty}|\tilde A|^2 du$ (black curve) and peak intensity (gray curve) of the propagating beam in (a) as a function of propagation distance $v$. (c) For $M=5$ and $\alpha=0$ (no Kerr nonlinearity), in green, amplitude $a_0(\infty)$ of the attracting NAB (solid curve), and its inward and outward H\"ankel amplitudes, $|a_{\rm in}(\infty)|$ and $|a_{\rm out}(\infty)|$ (dashed curves), as functions of the amplitude $a_0(0)$ of the input Airy beam. The same lines in black correspond to $\alpha=1$ (Kerr medium).}
\end{figure}

First, we address the case of an ideal Airy beam $\tilde A(u,0)=a_0Ai(u)$ at the entrance plane $v=0$ of the medium. This ideal situation allows us to distinguish the effects that are due to pure nonlinearities from those arising from apodization in finite-power Airy beams (which is known to cause also loss of contrast \cite{SIVI}). From the point of view of numerical simulations of Eq. (\ref{NLSEN1}), the strongly distorting effects caused by the finiteness of the computational box in unapodized Airy beams are eliminated by the procedure of replacing the propagated envelope $\tilde A$ at each numerical propagation step by the linearly propagated Airy beam in a narrow interval about the ends of the computational box. This procedure is justified because we know that nonlinear effects in the beam profile start, as for Bessel beams \cite{PORRAS3,PORRAS} at the Airy main maximum and propagate outwards. Of course, the numerical simulations are valid only up to the distance at which nonlinear effects reach the end of the computational box.

Each Airy beam is then numerically seen to undergo a transformation into a particular stationary NAB. An example of the dynamics toward the final stationary NAB is shown in Fig. \ref{Fig3}(a), where the input linear Airy with $a_0=3.5$ in a medium with $M=5$ and $\alpha=1$ transforms into the NAB defined by $a_0=4.13$. Accordingly, the total NLLs experienced during propagation and the peak intensity, shown in Fig. \ref{Fig3}(b), are seen to stabilize into a constant value equal to the NLLs $N_{-\infty}$ and peak intensity of the NAB with $a_0=4.13$. Figure \ref{Fig3}(c) summarizes the results of extensive numerical simulations. For each value of $a_0$ of the input Airy beam at $v=0$, named $a_0(0)$, the value of $a_0$ of the attracting NAB, named $a_0(\infty)$, is represented (solid curves). We also depict in Fig. \ref{Fig3}(c) the values of $|a_{\rm in}(\infty)|$ and $|a_{\rm out}(\infty)|$ of the NAB with $a_0(\infty)$ (dashed curves), as extracted from Fig. \ref{Fig1}(d). The crucial observation here is that, irrespective of the values of $M$ and $\alpha$, $|a_{\rm in}(\infty)|$ is given by the identity function $|a_{\rm in}(\infty)|=a_0(0)$. Since for the input linear Airy beam $a_{\rm in}(0)=a_{\rm out}(0)=a_0(0)$, we then infer that the amplitude of the inward H\"ankel component is preserved in the nonlinear propagation. This observation constitutes the response to the question of which particular NAB is formed given the input linear Airy beam. The final NAB is that whose inward H\"ankel amplitude $|a_{\rm in}|$ equals to the amplitude $a_0$ of the input Airy beam.

Here the question of whether a stationary NAB is always reached or not arises. Numerical simulations with apodized Airy beams in \cite{COUAIRON} indicate that a stationary state may not be reached because of an instability of the final NAB, this instability being observed as ``tangential" emissions during propagation. These emissions are incipiently seen in Fig. \ref{Fig3}(a), but they do not prevent our ideal, unapodized Airy beam from reaching the inward H\"ankel-preserving NAB. In additional simulations with increasing values of of $\alpha$ up to $\alpha=3$ (increasing relative strength of Kerr nonlinearity), these emissions become sharper. The final stationary NAB continues to be formed but much more slowly and therefore at much longer propagation distances. With ideal, unapodized Airy beams in media with $\alpha>3$, the numerical procedure explained above requires such a long propagation distance, and therefore wide transversal computational box, that it is difficult to draw any conclusion about whether the Airy beam enters into a unstable regime or converges eventually to the NAB. As pointed out in \cite{COUAIRON}, a stability analysis of NABs would help to elucidate this question. In this respect the situation is different with Bessel beams, where unstable nonlinear propagation regimes are clearly identified \cite{COUAIRON,POLESANA}.

\section{Dynamics of finite-power Airy beams in the nonlinear medium: abrupt and soft input}\label{APODIZED}

We consider now the nonlinear propagation Airy beams with finite power. As a representative case, we take the exponential-Airy beam \cite{SIVI}
\begin{equation}\label{LA}
\tilde A(u,v)=a_0 Ai(u+i\gamma v) e^{i\phi_L(u,v)} e^{\gamma(u\!-\! v^2/4)}e^{iv \gamma^2/2}
\end{equation}
generated, for example, by imprinting a cubic phase onto a Gaussian beam and using a 2--$f$ Fourier transform system \cite{LOTTI}. As is well-known, the transversal profile of this beam evolves from a nearly Gaussian profile at the far field ($v\rightarrow\pm\infty$) to the Airy-exponential profile $a_0Ai(u)\exp(au)$ at the ``waist" ($v=0$). We may then wonder whether its nonlinear propagation has something to do with that of the ideal, diffraction-free Airy beam, particularly when the entrance plane of the medium is placed at large negative $v$, where the profile approaches a Gaussian profile.

\begin{figure}[b]
\centering\includegraphics*[width=4.5cm]{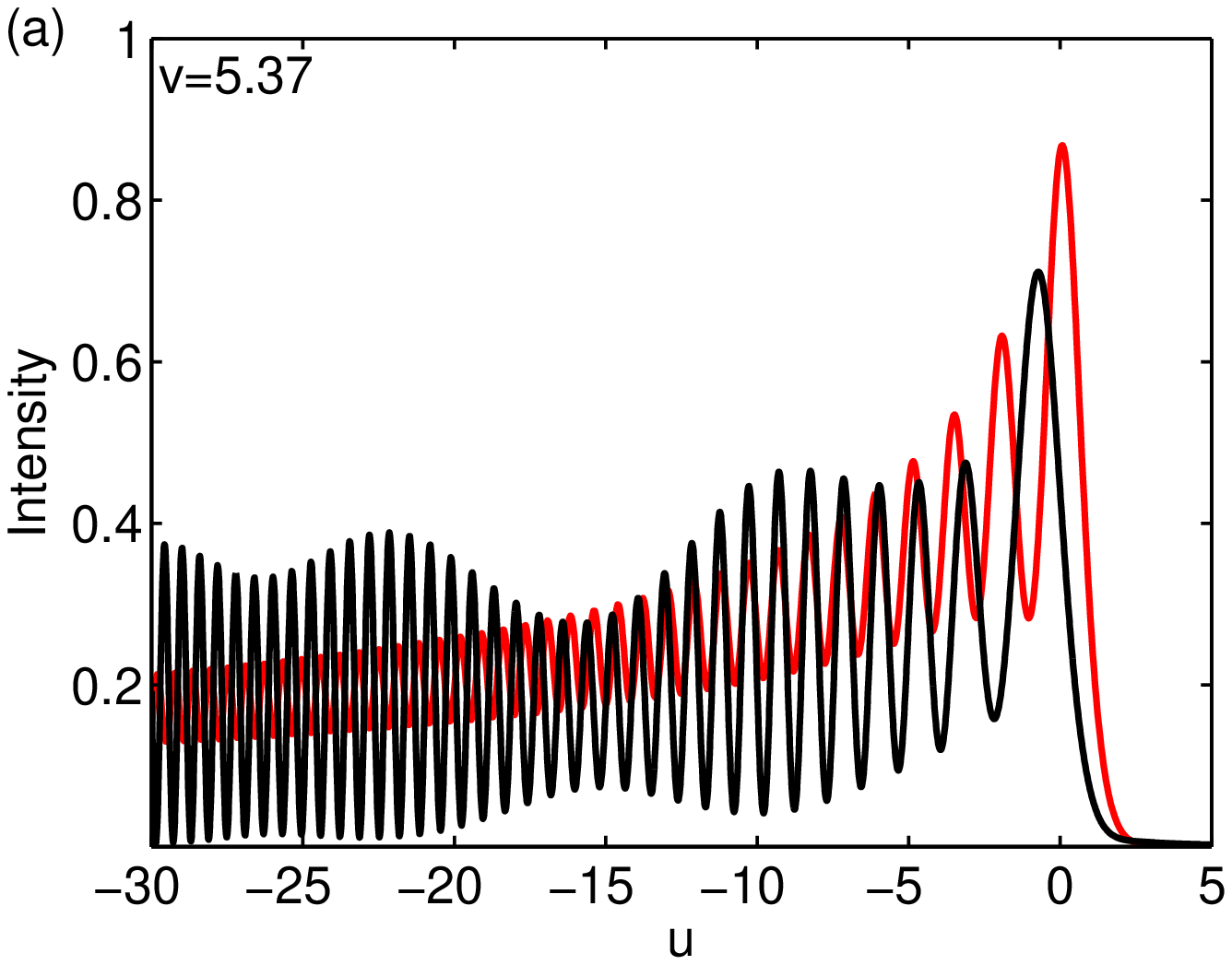}\includegraphics*[width=3.6cm]{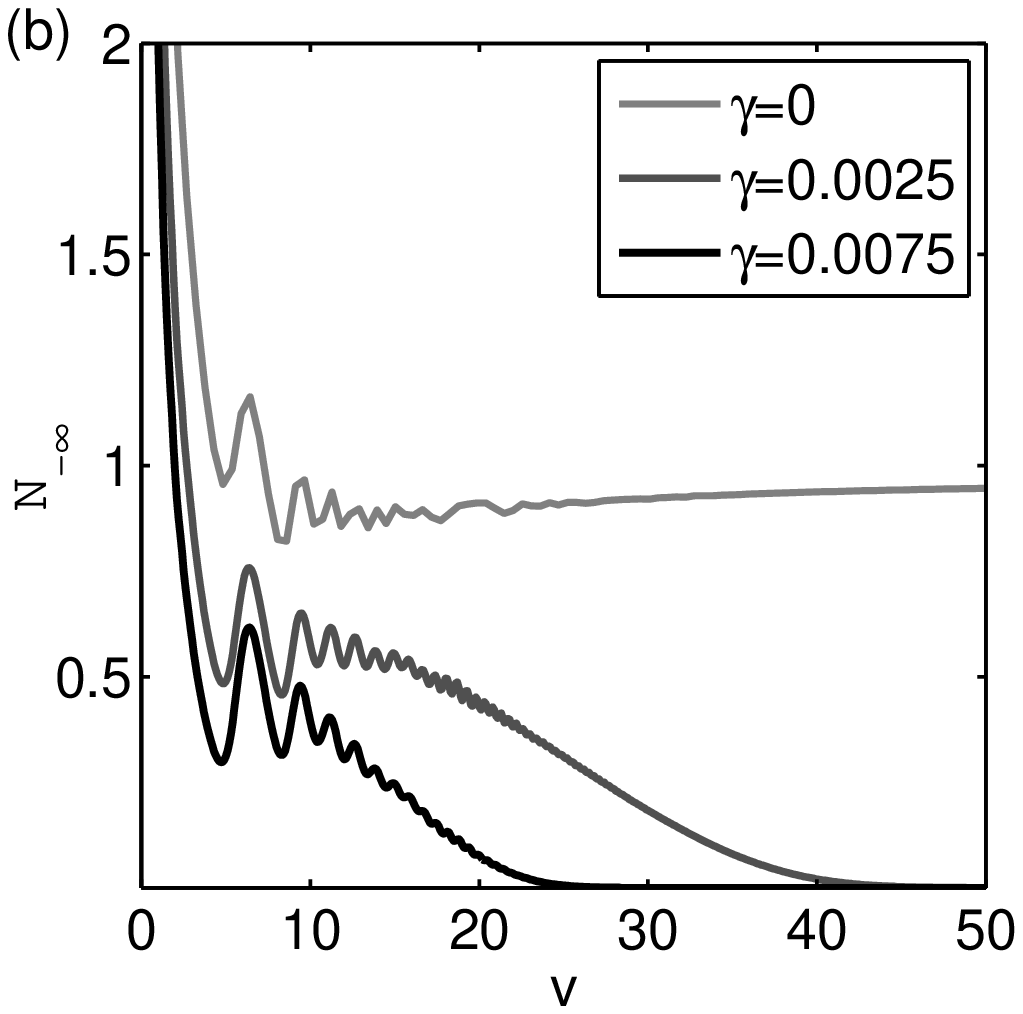}\includegraphics*[width=3.7cm]{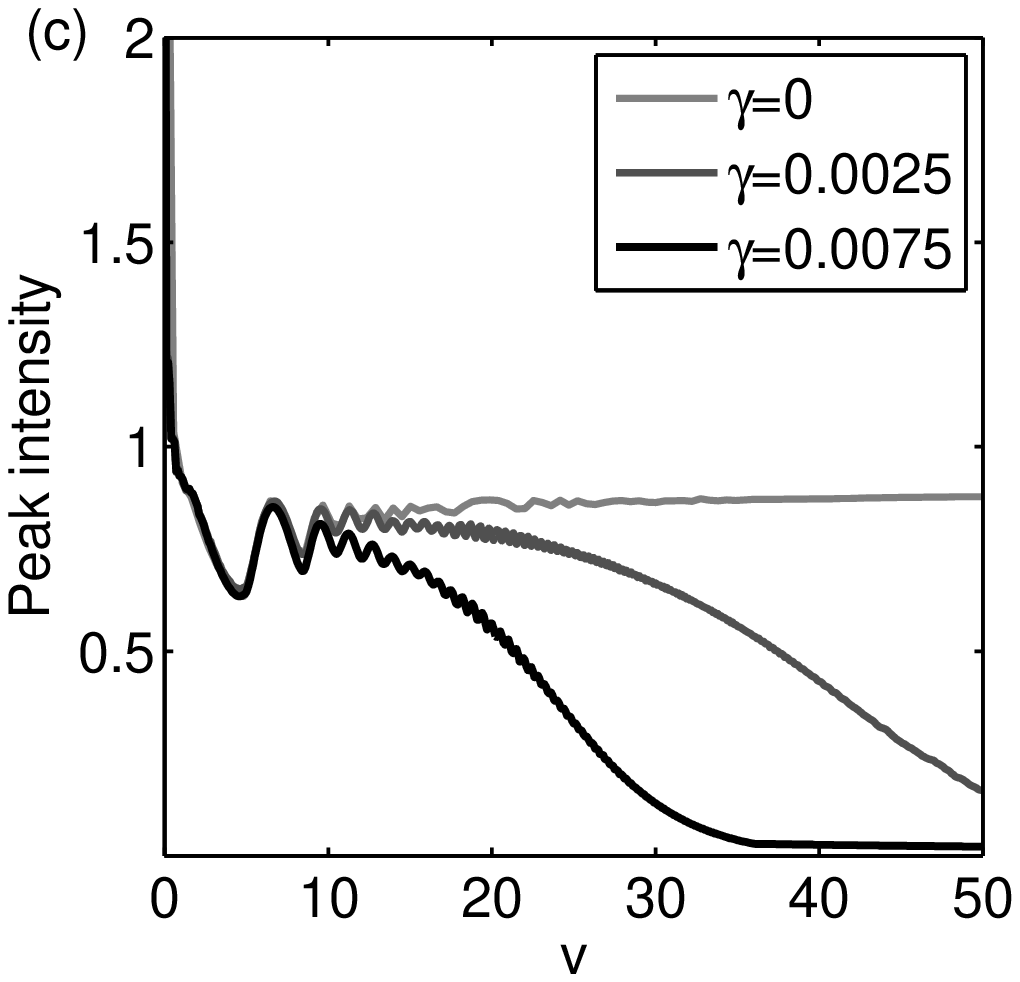}\\
\caption{\label{Fig4} For a medium with $M=5$ and $\alpha=1$ with entrance plane at $v=0$ (abrupt input): (a) \textcolor{blue}{Media˜2}: Change of the intensity profile of an initial exponential-Airy beam with $a_0=3.5$ and $\gamma=0.0025$ with propagation distance $v$. The red curve represents the ideal attracting NAB with $a_0=4.13$ ($|a_{\rm in}|=3.5$); (b) NLLs and (c) peak intensity versus propagation distance $v$ for $\gamma=0$, or ideal case (light gray), $\gamma=0.0025$ (dark gray), and $\gamma=0.0075$ (black).}
\end{figure}

Figure \ref{Fig4}(a) shows a typical example of the propagation dynamics of the exponential-Airy beam in Eq. (\ref{LA}) when the entrance plane of the medium coincides with the waist $v=0$. We will refer this configuration to as ``abrupt" input, since the intensity is maximum at the entrance plane and therefore the nonlinear effects are initially strong. A detailed comparison between Fig. \ref{Fig4}(a) and Fig. \ref{Fig3}(a) shows that during the first stages of propagation the apodized Airy beam follows the same dynamics of transformation towards the attracting NAB. It cannot be reached, however, because the finite power flowing towards positive $u$ is consumed before the slow process of NAB formation is concluded. The variation of the NLLs and of the peak intensity during propagation [Figs. \ref{Fig4}(b) and \ref{Fig4}(c)] are seen to reproduce initially those of the ideal case, but at lower levels of NLLs and intensities that further decrease down to zero.

\begin{figure}
\centering\includegraphics*[width=4.5cm]{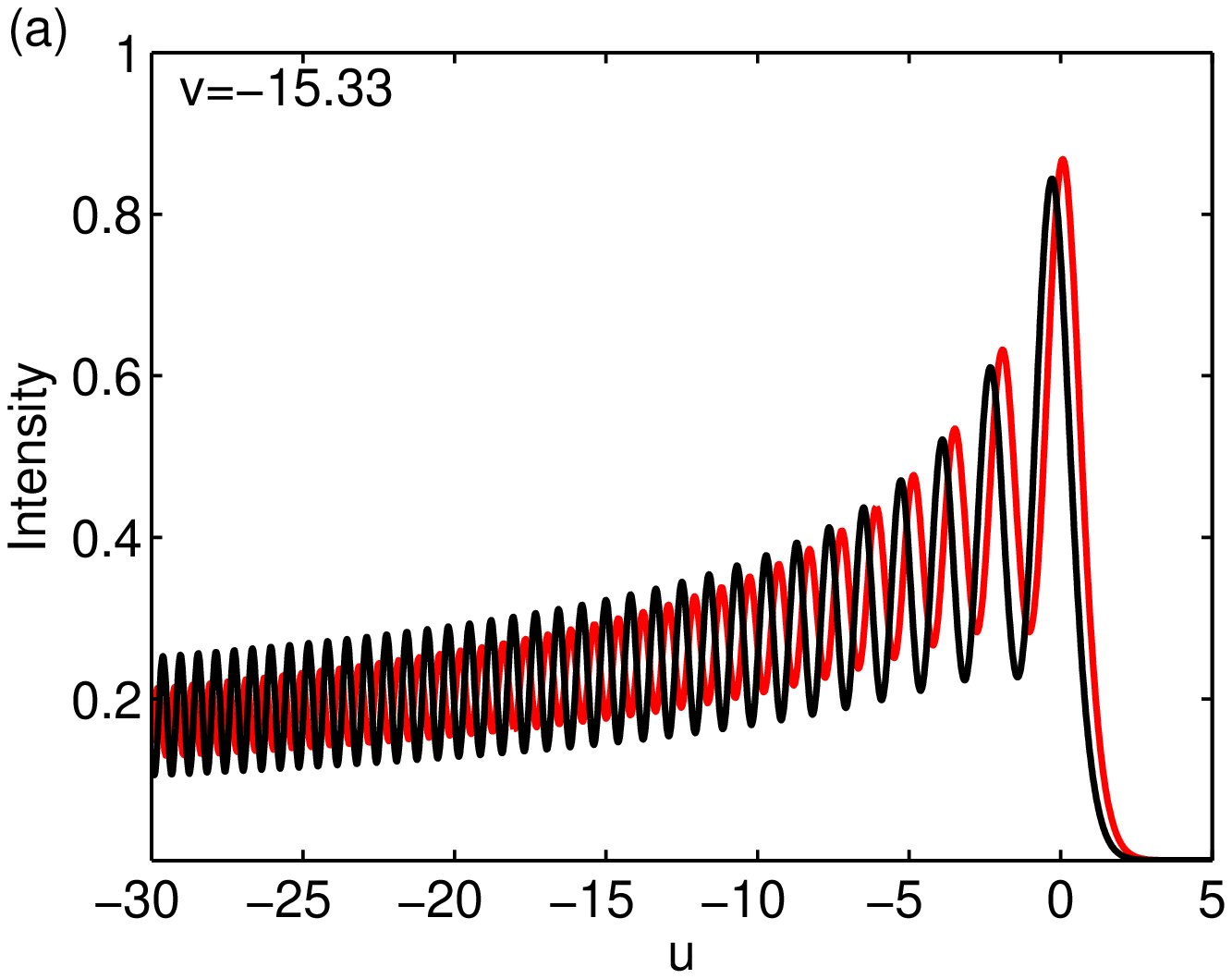}\includegraphics*[width=4.5cm]{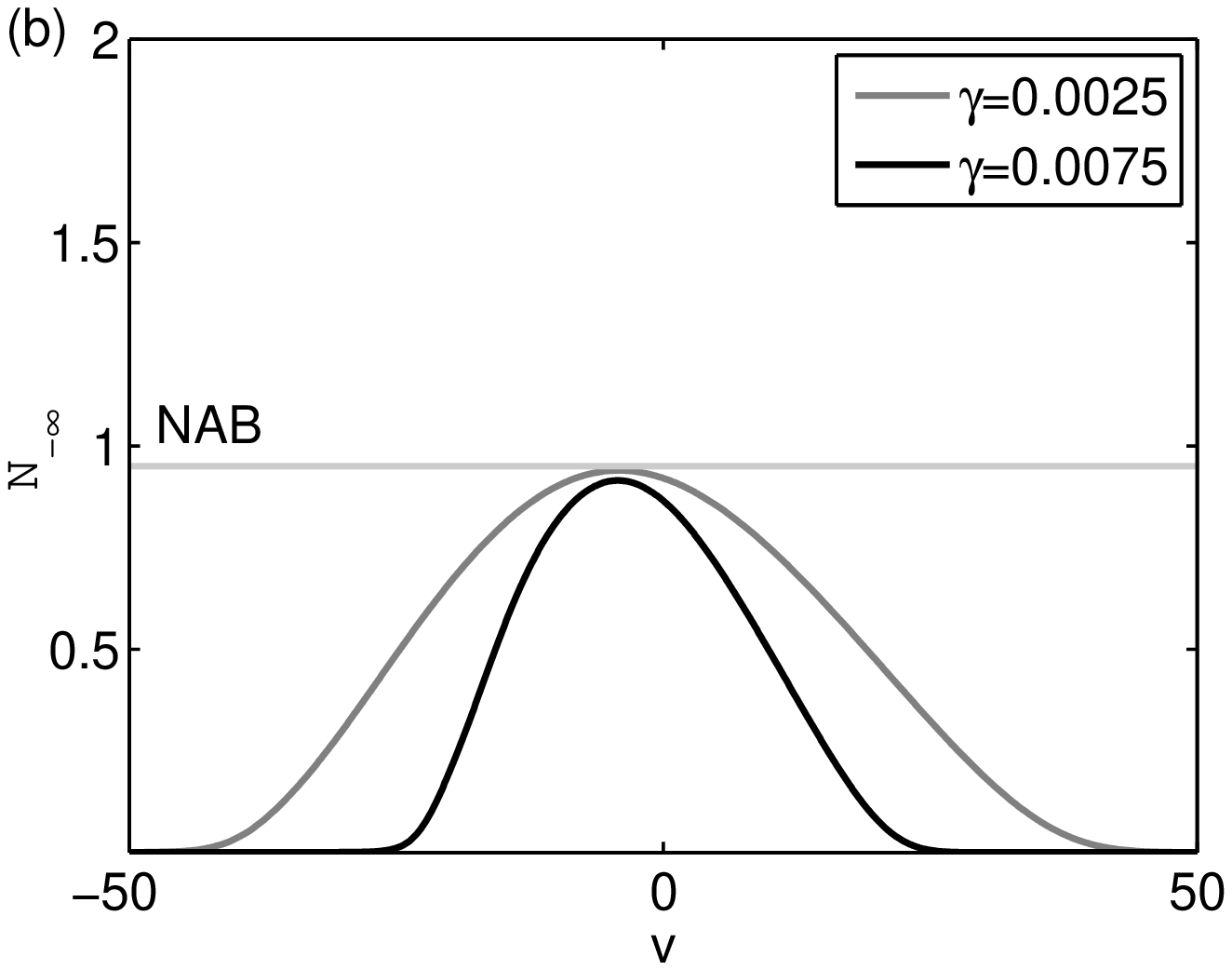}\includegraphics*[width=4.5cm]{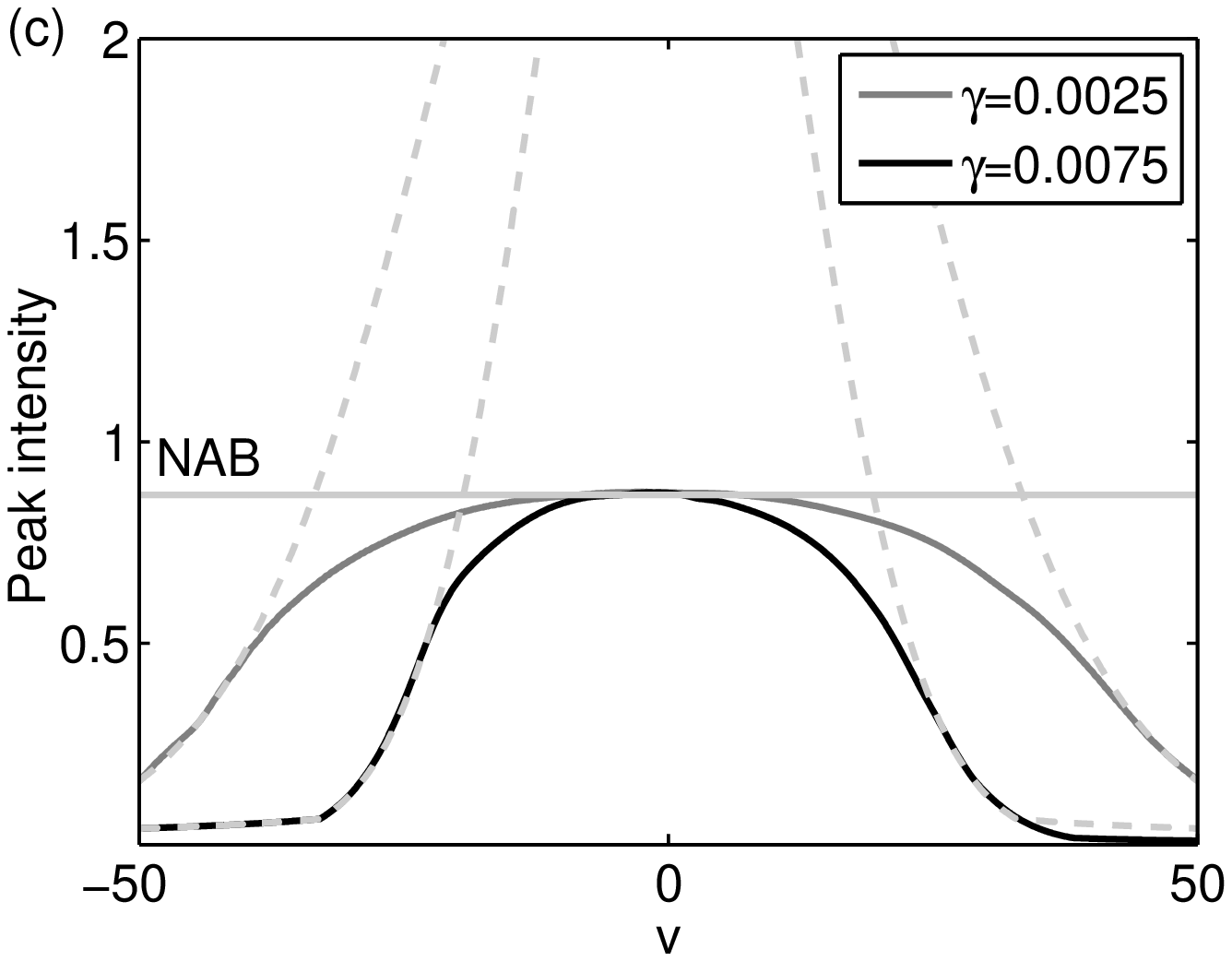}\\
\caption{\label{Fig5} For $M=5$ and $\alpha=1$ and entrance plane at $v=-50$ (soft input): (a) \textcolor{blue}{Media˜3}: Change of the intensity profile of an initial exponential-Airy beam with $a_0=3.5$ and $\gamma=0.0025$ with propagation distance $v$. The red curve represents the ideal attracting NAB with $a_0=4.13$ ($|a_{\rm in}|=3.5$); (b) NLLs and (c) peak intensity versus propagation distance $v$ for $\gamma=0.0025$ (dark gray) and $\gamma=0.0075$ (black). The horizontal lines represent the NLLs and the peak intensity of the attracting NAB. The dashed gray curves in (c) represent the peak intensity of exponential-Airy beams propagating linearly.}
\end{figure}

Figure \ref{Fig5} refers to the propagation of the same exponential-Airy beam as in Fig. \ref{Fig4} except that the entrance plane in the medium is at $v<0$ sufficiently  before the waist for the beam to be widespread at low intensity levels ($|\tilde A|^2\ll 1$) and for NLLs to be therefore  initially negligible. The dynamics observed in Fig. \ref{Fig5}(a) is independent of the particular choice of the entrance plane $v<0$ provided that this "soft" input condition is satisfied. The remarkable fact is that, in spite of the completely different behavior of the transversal profile, the same law of attraction by the stationary NAB that preserves the amplitude of the inward H\"ankel component presides the dynamics. In contrast to abrupt input, the attracting NAB is reached in a broad region about the focus, a transversal region whose width depends on the apodization strength. Figures \ref{Fig5}(b) and \ref{Fig5}(c) show that the NLLs and peak intensity do not present the ripples of the abrupt case, but reach smoothly values close to the NLLs and peak intensity of the attracting NAB.

Thus, in spite of the variety of situations, the propagation of finite-power Airy beams is primarily governed by the transformation towards the same attracting NAB as in the ideal case. In turn, this NAB is solely determined by the parameter $a_0$ of the input Airy beam, or what amounts the same thing, by the maximum intensity $\simeq 0.287 a_0^2$ of the input linear Airy beam (or the maximum intensity that would reach at the focus for soft input).

\begin{figure}
\centering\includegraphics*[width=6cm]{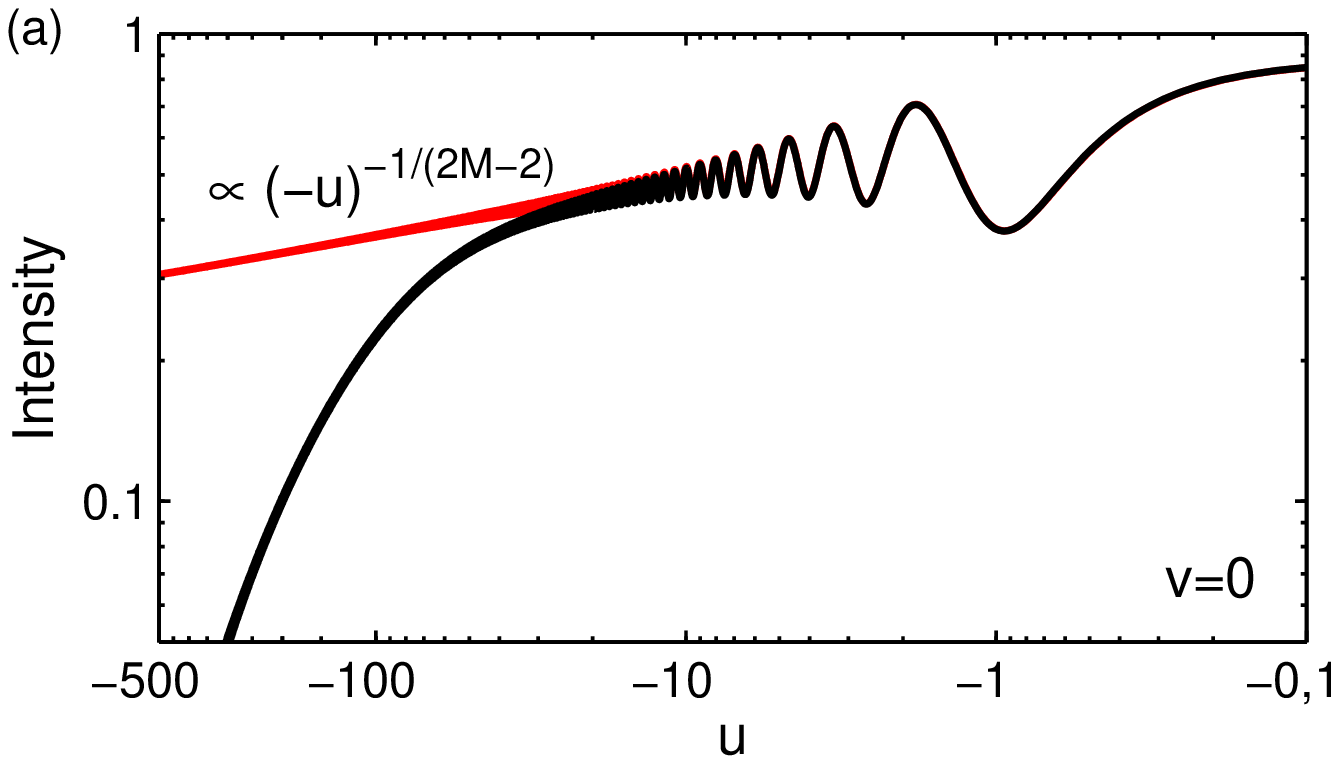}\includegraphics*[width=7cm]{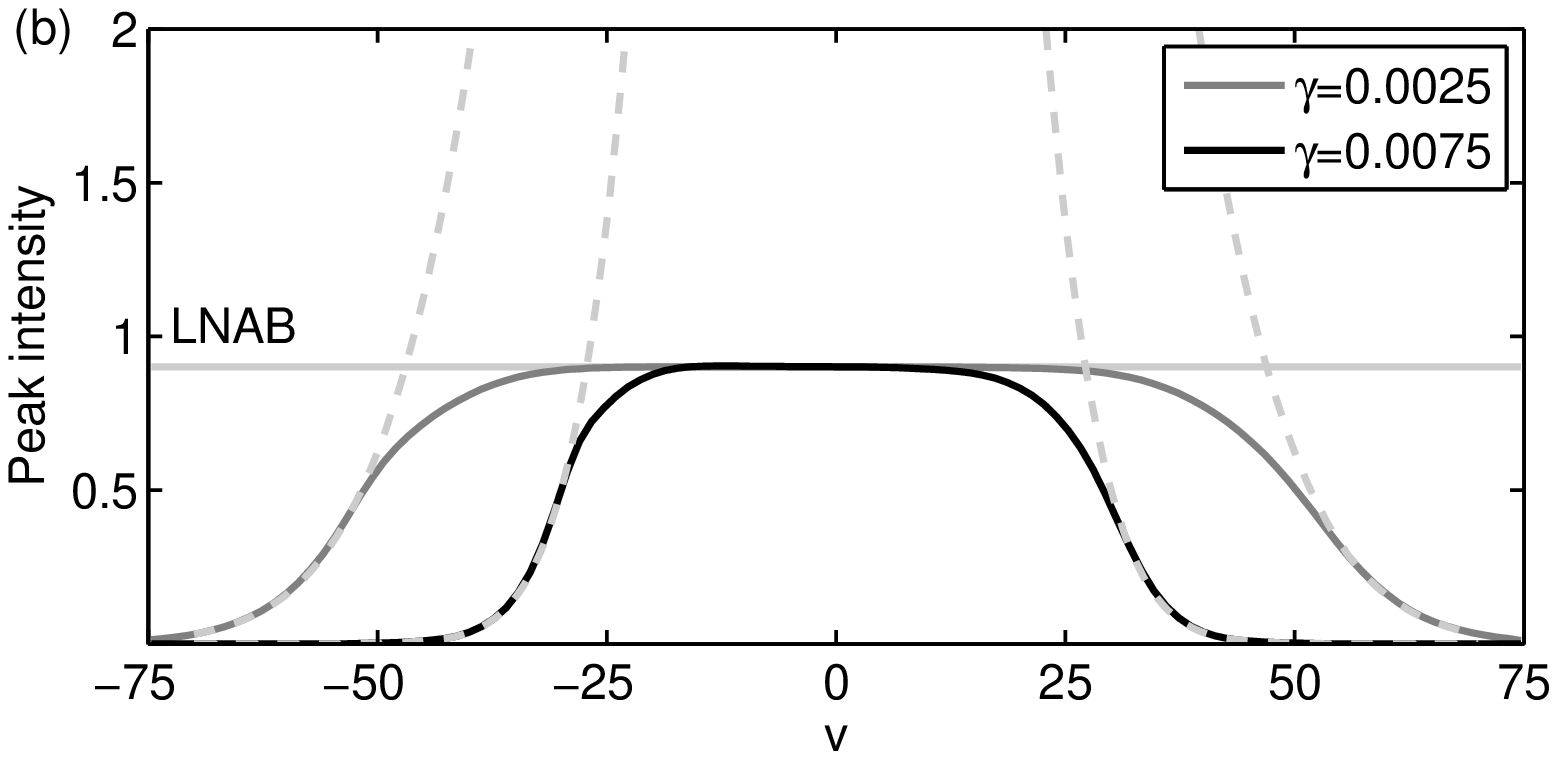}
\caption{\label{Fig6} For $M=5$ and $\alpha=1$ and entrance plane at $v=-75$ (soft input): (a) intensity profile at the focus $v=0$ of an input exponential-Airy beam with $a_0=7.0$ and $\gamma=0.0025$ (black curve), and intensity profile of the LNAB in this medium (red curve). (b) Peak intensity versus propagation distance for $\gamma=0.0025$ (dark gray) and $\gamma=0.0075$ (black). The horizontal line represents the peak intensity of the LNAB. The dashed gray curves represent the peak intensity of exponential-Airy beams propagating linearly.}
\end{figure}

Consider finally the limiting strong nonlinear regime of increasing $a_0$. The attracting NABs are then characterized by $a_0$ approaching $a_{0\rm max}$, and, as seen in Sec. \ref{ASYMPINFINITE}, these NABs tend to match the LNAB with $a_{0,\rm max}$ in a transversal region expanding towards $u\rightarrow -\infty$. From a practical point of view, any exponential-Airy beam smoothly launched in the medium will result in the partial formation the LNAB irrespective of the large value of $a_0$, as illustrated in the example of Fig. \ref{Fig6}(a), since apodization prevents from the observation of any (ever further) H\"ankel tails.

As seen in Figs. \ref{Fig5}(c) and \ref{Fig6}(b) for the peak intensity under soft input conditions, the formed NAB or LNAB persists for a propagation distance that increases with $a_0$ and weakening apodization. Observing in these figures that the peak intensity rises initially, and decreases after the plateau region of intensity about unity, approximately as that of the corresponding linear Airy beam (gray dashed curves) $\simeq 0.287 a_0^2\exp(-\gamma v^2/2)$, we readily obtain the estimation
\begin{equation}
\Delta v\sim 2\sqrt{2\ln(0.287 a_0^2)/\gamma}
\end{equation}
(provided that $0.287 a_0^2>1$) for the distance of sustained peak intensity. The length of this region can be made arbitrarily large not only by weakening the apodization, as for linear Airy beams, but also by increasing the intensity of the input Airy beam.

\begin{figure}
\centering\includegraphics*[width=6cm]{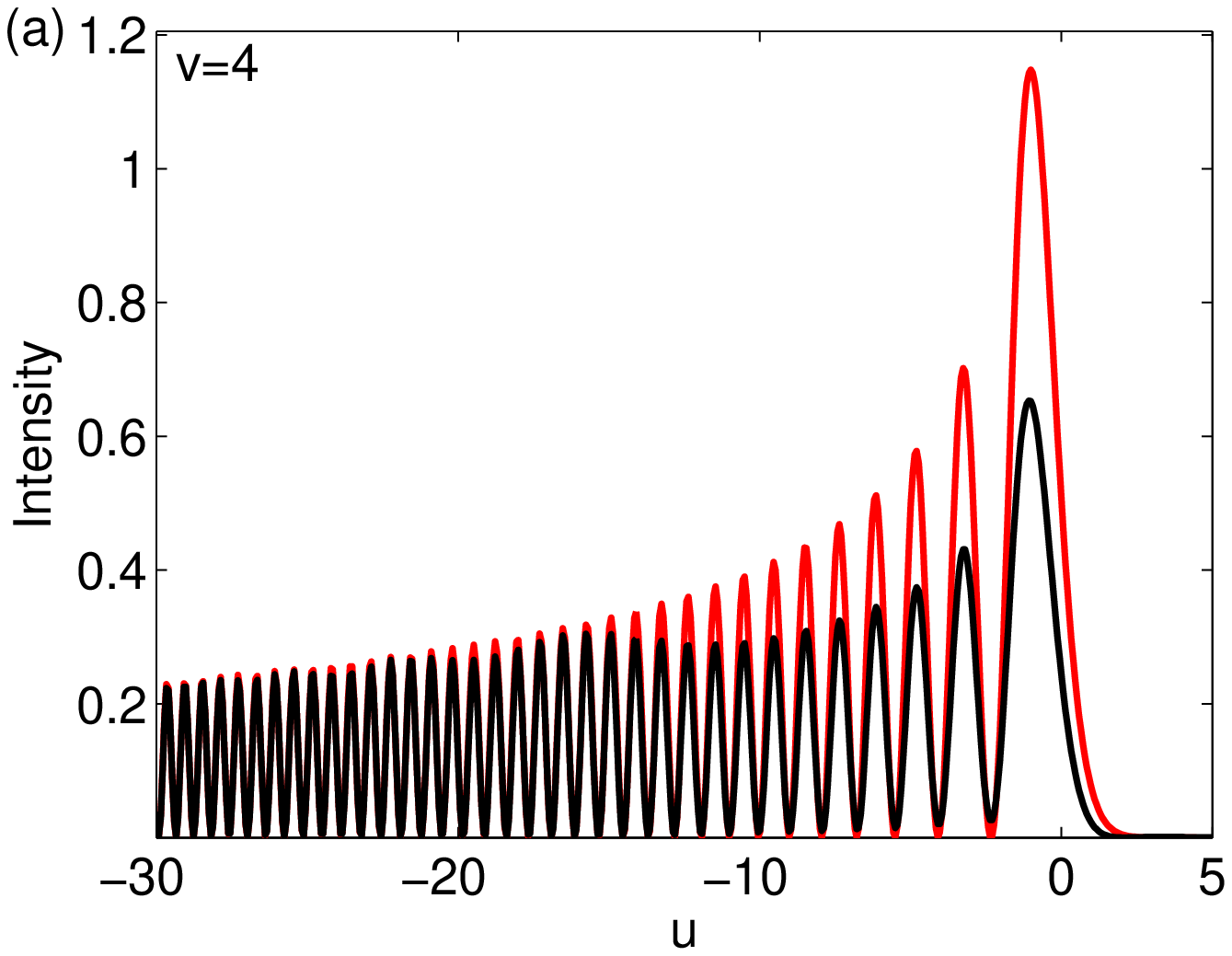}\includegraphics*[width=6cm]{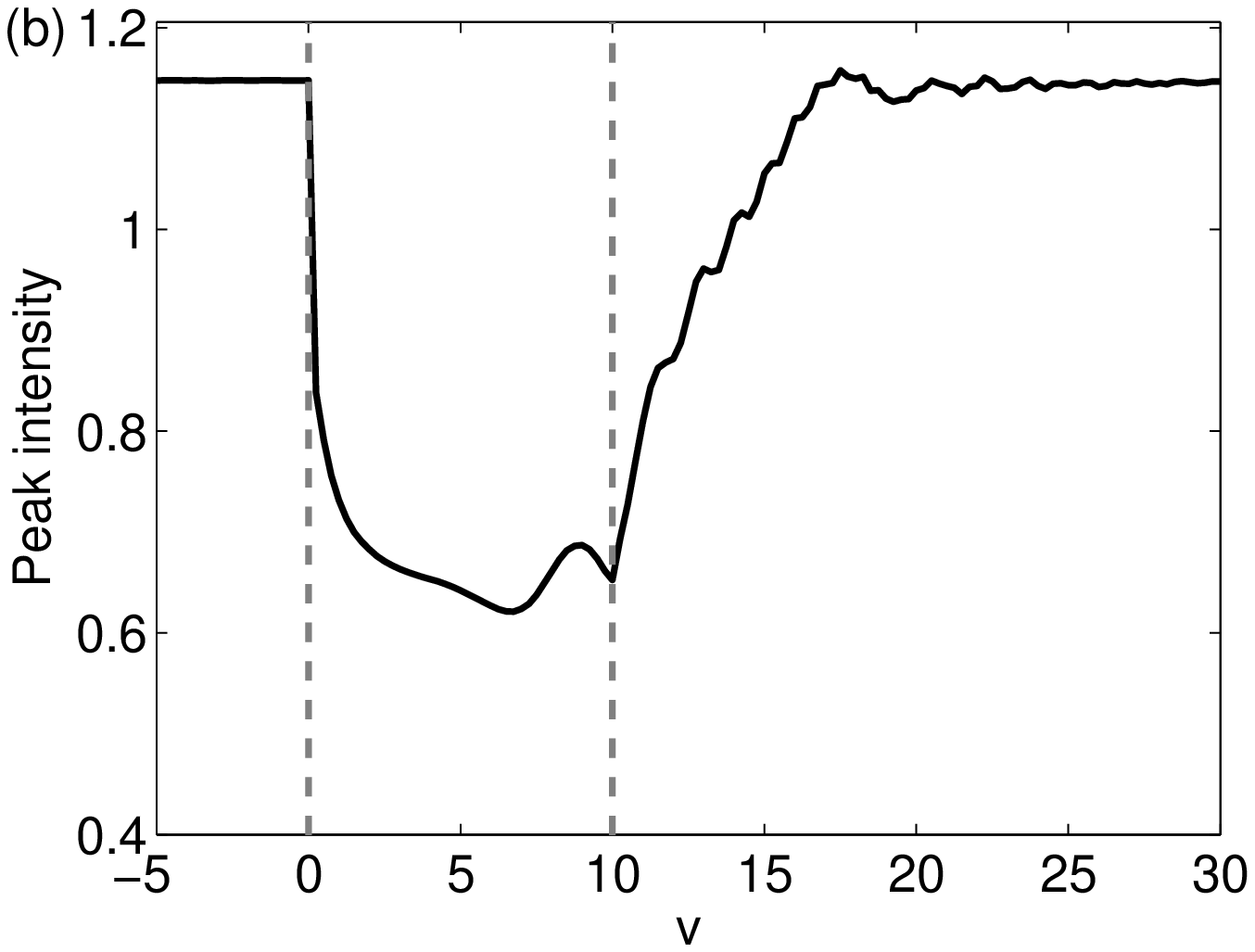}
\caption{\label{Fig7} Dynamics of a linear Airy beam with $a_0=2$ entering a nonlinear medium with $M=5$ and $\alpha=0$ at $v=0$ and exiting it at  $v=10$. (a) \textcolor{blue}{Media˜4}: Change of the intensity profile with propagation distance. The red curve represents the initial lineal Airy. (b) Peak intensity versus propagation distance. The dashed vertical lines represent the region filled by the nonlinear medium.}
\end{figure}

\section{Conclusions}

We have described the nonlinear propagation of Airy beams in media with significant NLLs due to multiphoton absorption. Though we have restricted the expressions and examples to media with self-focusing Kerr nonlinearity, similar description is expected to hold for other dispersive nonlinearities, or with the inclusion of ionization and plasma effects, as is also the case with Bessel beams \cite{JUKNA}.

As expected from \cite{LOTTI}, NABs play an essential role in the nonlinear propagation of Airy beams. We have provided with a full asymptotic analysis of NABs. This provides the basis for understanding the nonlinear dynamics of the initial Airy beam. This asymptotic analysis itself shows that there exist two types of NABs: the NABs described in \cite{LOTTI} with asymptotic linear, unbalanced H\"ankel decay as $1/(-u)^{1/2}$ in intensity, and the limiting NAB wich is fully nonlinear and infinitely lossy, decaying as slowly as $1/(-u)^{1/(2M-2)}$.

Our numerical simulations with initial ideal or finite-power versions of Airy beams reveals that a stationary NAB tends always to be formed, and unveils a conservation law that determines the particular NAB that attracts the input Airy beam, namely, the NAB that preserves the inward H\"ankel component of the input Airy beam. This conservation law allow to understand the seemingly different behaviors that are observed in the dynamics with input ideal or real Airy beams.

This law appears to hold under much more general conditions, and therefore deserves further consideration in related problems, as in the nonlinear propagation of Bessel beams. In Figs. \ref{Fig7}(a) and \ref{Fig7}(b), for example, an Airy beam enters and exits a medium with NLLs of finite length. The conservation law implies what we can call a {\em dissipative but reversible dynamics}: The Airy beam is attenuated to a NAB, but once the beam leaves the medium, the initial Airy beam is restored, as required by the conservation of the inward H\"ankel component. It is interesting to compare this situation with the counterintuitive situation of the damped oscillations by friction whose initial amplitude is recovered after friction is switched off. This counterintuitive behavior, closely related to the self-healing property of Airy beams \cite{BROKY}, is possible because the Airy beam can be regarded to be an open system that interchanges continuously energy with an energy reservoir.

\section*{Acknowledgments}

M.A.P. acknowledges support from Projects No. MTM2012-39101-C02-01 and No. FIS2013-41709-P of the Spanish Ministerio de Econom\'{\i}a y Competitividad, and K. Z. N\'obrega acknowledges CNPq/PDE and IFMA for financial support.


\begin{thebibliography}{99}
\bibitem{LOTTI2} Note that the signs in the exponentials are opposite in \cite{LOTTI}.
\end{thebibliography}
\end{document}